\documentclass[12pt,a4paper]{article}
\usepackage{jheppub}
\usepackage{dsfont}
\usepackage{amssymb, amsmath}
\usepackage{epsfig}
\usepackage{epstopdf}
\usepackage{latexsym}
\usepackage{graphicx}
\usepackage{subfigure}
\usepackage{booktabs}
\usepackage{bbm}
\pdfoutput=1
\makeatletter
\def\@fpheader{\relax}
\makeatother

\def\E{{\cal E}}

\def\R{{\cal R}}

\def\E{{\cal E}}
\def\O{{\cal O}}
\def\D{{\cal D}}

\def\cF{{\cal F}}
\def\cM{{\cal M}}
\def\Z{{\cal Z}}
\newcommand{\be}{\begin{equation}}
\newcommand{\ee}{\end{equation}}
\newcommand{\bea}{\begin{eqnarray}}
\newcommand{\eea}{\end{eqnarray}}
\newcommand{\bear}{\begin{eqnarray}}
\newcommand{\eear}{\end{eqnarray}}
\newcommand{\beas}{\begin{eqnarray*}}

\newcommand{\eeas}{\end{eqnarray*}}
\newcommand{\ba}{\begin{array}}
\newcommand{\ea}{\end{array}}

\title{Fast Scrambling under an RG-flow}
\author{Arnab Kundu$^{a}$}
\affiliation{$^a$Theory Division, Saha Institute of Nuclear Physics, HBNI, 1/AF Bidhannagar, Kolkata- 700064, India.}
\emailAdd{arnab.kundu [at] saha.ac.in}

\abstract {The notion of information scrambling is tied to the long-time behaviour of a system and therefore is related to its infra-red dynamics. In fast scramblers, information spreads as a logarithmic function of the number of degrees of freedom. Ordinarily, under an RG-flow, scrambling will become faster in the IR since many degrees of freedom are integrated out, as a consequence of the c-theorem. In this article, we consider a class of Holographic quantum field theories (QFT), which are strongly coupled large $N$ gauge theories with large number of adjoint and fundamental matter, in which scrambling slows down in the IR. This happens since more degrees of freedom are inserted in the IR, compared to the UV, in a precise sense. For generic large $N$ gauge theories, we also explore general, perturbative flow features of the corresponding Lyapunov exponent, based on the Callan-Symanzik equation.}

\begin{document}

\maketitle
\flushbottom

\section{Introduction}

Given a quantum mechanical system and a particular state, the time-scale involved for a localized information to spread across the entire system is set by the scrambling time. The physics of information and its propagation has become one of the most explored areas of theoretical physics in recent times. It turns out that scrambling time has a particularly illuminating role for quantum systems that possess a well-defined semi-classical limit.\footnote{Thus, for most spin systems, including the Ising model, although it can be defined, scrambling time may not set a particularly specific scale in the system.} There exists multiple definitions of scrambling time, in this article, we will consider the most widely used definition in terms of out-of-time-order correlators (OTOC), in the context of the fast scramblers\cite{Hayden:2007cs, Sekino:2008he, Lashkari:2011yi, Yoshida:2017non}.

In this regard, a particularly interesting class of systems are the ones with a large number of degrees of freedom, $N$, in which the inverse number provides a natural small parameter in the system, using which one can define a semi-classical limit. Typically, in standard quantum mechanical systems, the corresponding scrambling time scales as a polynomial of $N$. For strongly coupled systems, with a holographic dual description, however, this scaling becomes logarithmic $\sim \log N$, and in \cite{Sekino:2008he}, it was conjectured to be the fastest scrambler in Nature.

Holographic systems, by virtue of universality near a black hole horizon, naturally realize the fast scrambling feature in terms of the dual quantum field theory degrees of freedom. Typically, for an SU$(N)$ gauge theory at the fixed point, the corresponding time-scale is given by $t_{\rm scr} \sim \log N $, which is essentially determined by the ten-dimensional Newton's constant, in the gravitational background. This description is valid when the Newton's constant is small (measured in the energy-scale of the physical process), equivalently when the number of degree of freedom in the dual QFT is large. It is, however, easy to observe that, even at the fixed point, not all degree of freedom in will scramble as $\log N$.

A simple explicit example is to consider, {\it e.g.}~an AdS-Schwarzschild background and study the dynamics of a probe string which is stretched along the radial direction of AdS. In the dual QFT, this is equivalent to introducing a fundamental defect matter ({\it e.g.}~like a quark) in an adjoint matter bath. The modes on this string come equipped with the string tension and translated into the language above, one obtains the scrambling of the open string modes to scale as: $t_{\rm scr} \sim \log \lambda$,\footnote{This is not strictly true, but qualitatively so. We will discuss the precise form momentarily.} where $\lambda$ is the 't Hooft coupling. In the strongly coupled limit, this still allows for a well-defined separation of time-scales between {\it e.g.}~dissipation time and scrambling time.

Within the purview of holographic dual, one considers large $N$ and large $\lambda$ and {\it a priori} it is not obvious whether {\it all possible} degrees of freedom in the system will scramble at scales determined by $\log N$ or $\log \lambda$. For example, functionally, a dependence of $\log \left( N / \lambda\right) $ is also possible. While, one can still tune the ratio $(N/\lambda)$ as one pleases, to take a large enough value, it is not necessarily a large number from the QFT-perspective. Demanding a geometric dual, however, imposes constraints on the curvature and the dilaton field, which ensure that such a ratio of large numbers will still remain large in the semi-classical regime. Thus, an $\O(1)$ scrambling time is naturally associated with either higher curvature corrections, or string loop corrections.

This article is divided into two main parts. In the first part, we improvise based on the standard Callan-Symanzik equation to infer statements on how the Lyapunov exponent should flow near a vanishing beta function. This analyses comes with certain technical and conceptual assumptions, which we only impose and not attempt to prove from a first principle calculation. This is to explore how much mileage a naive approach might yield, and we leave a more systematic exploration for future. For example, we impose an {\it ad hoc} relation between energy and time, which can be replaced by a general Kibble-Zurek type scaling, see {\it e.g.}~\cite{Das:2016eao}. Nevertheless, the Callan-Symanzik equation yields interesting results that warrantees further exploration.

In the second part, we consider explicit and representative examples within the purview of Holographic renormalization group flow, in which the corresponding scrambling time also flows. The corresponding QFTs are standard SU$(N)$ Yang-Mills system, with $N_q$ density of fundamental matter. We will pose the physical question in the following form: Given a UV-CFT (which could simply be a Gaussian fixed point), where we fix all scales  of measurement, we will ask questions about the thermal state of the system at different energy scales. Physically the picture is simple: Suppose, in a room full of ordinary air molecules, we ask questions related to thermal physics resulting from the collision of high energy protons. In this case, it is expected that a quark-gluon plasma will describe the corresponding thermal physics. On the other hand, in the same room, if we burst some fire-crackers, the resulting thermal physics will very well be described by an ordinary thermal gas. Clearly, these two thermal physics, which are intrinsically connected to the energy-scale of the collision process (protons colliding, or bursting of fire-crackers) are quite different.

Similarly, given a UV-CFT, a thermal state can be obtained at UV, which will be described by the UV degrees of freedom. On the other hand, provided the CFT allows for a non-trivial RG-flow to an IR-CFT, a different thermal description can also be obtained at the IR, in terms of very different degrees of freedom. Our measurement apparatus is set up at the UV, and we will analyze what it can tell us about scrambling of various degrees of freedom, both at the UV and at the IR. This approach is particularly suited for Holography, where one sets up the physical question in precisely the manner above, by fixing scales and measuring data at the UV conformal boundary.

Moreover, as we will explicitly see, Holography can repackage the number of degrees of freedom under an RG-flow in sufficiently non-trivial manner such that non-trivial combinations of large number of degrees of freedom and the 't Hooft coupling can emerge in the IR. Thus, naively, though one would expect the scrambling time to decrease under an RG-flow --- since now information has to spread across a smaller number of degrees of freedom --- it depends on the particular operator. This occurs mainly due to different probes coupling differently in the system, and also since the RG-flow is triggered by inserting new degrees of freedom at the UV, which is then non-trivially entangled in the IR. In fact, on the class of examples, we observe that scrambling time generically increases in the IR. Note, however, that a strict monotonicity does not seem assured from the dynamical equations of gravity, upon using reasonable energy conditions on the matter sourcing the geometry.

This article is divided in the following parts: In section $2$, we explore how much information the Callan-Symanzik equation can yield on out-of-time-order correlators (OTOCs) from which one extracts the corresponding ergodic data. After discussing various possibilities we consider the Holographic examples in the next sections. Finally, some technical details are assorted in an Appendix.

\section{Scrambling and RG: General Results}

In this section we present generic results that one can obtain by using elementary RG-ideas and equations. First of all, since we will be considering the $4$-point OTOCs, it makes sense to resort to the equation that correlators satisfy under an RG-flow: the Callan-Symanzik equation. Consider a generic $4$-point correlator:
\begin{eqnarray}
G^{(4)}= \left \langle \phi(x_1) \phi(x_2) \phi(x_3) \phi(x_4) \right \rangle = z^{-4/2} \left \langle \phi_0(x_1) \phi_0(x_2) \phi_0(x_3) \phi_0(x_4) \right \rangle = z^{-2} G_0^{(4)}  \ , \nonumber \\
\end{eqnarray}
where $\phi_0$ is the bare field and $z$ is the wavefunction renormalization constant: $\phi = z^{-1/2} \phi_0$. The correlator $G^{(4)}$ is a function of renormalized quantities, {\it e.g.}~fields, coupling and mass; on the other hand, $G_0^{(4)}$ depends only on the bare quantities. Therefore, by construction:
\begin{eqnarray}
\mu\frac{d G_0^{(4)}}{d\mu} = 0 \ ,
\end{eqnarray}
where $\mu$ is the corresponding energy-scale. The equation above can be rewritten in terms of the physical correlator, which yields the Callan-Symanzik equation:
\begin{eqnarray}
\left( \mu \frac{\partial}{\partial \mu} + \beta \frac{\partial}{\partial g} + 4 \gamma \right) G^{(4)} = 0 \ , \label{callansymanzik}
\end{eqnarray}
where $g(\mu)$ is the (dimensionless) coupling constant, $\beta$ is the corresponding beta-function and $\gamma$ is the anomalous dimension of the field $\phi$. Note that, in deriving the above equation, we need not assume anything about time-ordering of the operators inside the correlator.

We will momentarily use the $4$-point OTOC, of the following form:
\begin{eqnarray}
G_{\rm norm}^{(4)} = \frac{\left \langle \phi(0) \phi(t) \phi(0) \phi(t) \right \rangle}{\left \langle \phi(0) \phi(0) \right \rangle \left \langle \phi(t) \phi(t) \right \rangle} \ ,
\end{eqnarray}
where we have normalized the four-point function by a product of two two-point functions. On scaling arguments, therefore, $G_{\rm norm}^{(4)}$ will receive no factor of wave-function renormalization $z$, hence the corresponding Callan-Symanzik equation for the normalized correlator will be independent of the anomalous dimension:
\begin{eqnarray}
\left( \mu \frac{\partial}{\partial \mu} + \beta \frac{\partial}{\partial g}  \right) G_{\rm norm}^{(4)} = 0 \ , \label{callansymanzik2}
\end{eqnarray}
which is the equation we will deal with.\footnote{Subsequently, all ergodic features of the correlators will be independent of the critical exponents of a putative fixed point, since the critical exponents are determined by the anomalous dimension.} Note that, we assume that, even in the dynamical context the Callan-Symanzik equation receives no additional renormalization constant and remains the same. This may not hold true in general.\footnote{See {\it e.g.}~\cite{Goykhman:2018iaz} for a detailed analysis of a global quench from an RG-perspective, in the $\phi^4$-theory. Furthermore, see also \cite{DeSarker:2013abc}. We thank Diptarka Das for emphasizing this point as well as the first reference. We also thank Krishnendu Sengupta for pointing a conversation in related matters and for pointing out the second reference.}

Now, let us consider an ansatz for the $4$-point normalized OTOC. Motivated by ergodic quantum systems with a well-defined semi-classical limit, in the large time limit is expected to behave as:
\begin{eqnarray}
G_{\rm norm}^{(4)} = 1 - \Gamma\left[ N(\mu), g(\mu) \right] {\rm exp}\left[ \lambda_L\left( g(\mu), N(\mu), \mu \right) t \right] \ , \label{ergodic}
\end{eqnarray}
where $N(\mu)$ estimates the total number of degrees of freedom at the energy-scale $\mu$ (which is a dimensionful quantity), $g(\mu)$ is the dimensionless coupling constant and $\Gamma$ is a functional of these two basic parameters of the system. We are restricting our discussion, for the time-being, to only one coupling. Note that, the ansatz in (\ref{ergodic}) is not an entire function for the normalized correlator, instead it is only valid within the window: $t_d < t < t_{\rm scr}$, where $t_d$ denotes the dissipation time and $t_{\rm scr}$ denotes the scrambling time. Now, to ensure a semi-classical limit, we insist:
\begin{eqnarray}
 \Gamma\left[ N(\mu), g(\mu) \right] \ll 1 \ . \label{sccond}
\end{eqnarray}
Note that, when $N(\mu) \gg 1$, this is ensured by choosing $\Gamma \sim 1/N$,\footnote{Any non-trivial positive power of $N(\mu)$ can also appear here, but in that case, we can define the corresponding expression as $N(\mu)$ itself, since any positive power of a large number is also a large number.} irrespective of the functional dependence of $\Gamma$ on $g(\mu)$, assuming that $g(\mu)$ does not scale with $N(\mu)$.\footnote{Note that, this is a widely made assumption. We thank Diptarka Das for discussion about this.} Namely, we can make a simple choice:
\begin{eqnarray}
 \Gamma\left[ N(\mu), g(\mu) \right] = \frac{g(\mu)^\alpha}{N(\mu)} \ , \quad \alpha \in {\mathbb R} \quad \& \quad g\not = 0 \ .
\end{eqnarray}
On the other hand, if $N(\mu) \sim \O(1)$, then (\ref{sccond}) demands:\footnote{Once again, we are only choosing a simple representative case.}
\begin{eqnarray}
&& \Gamma\left[ N(\mu), g(\mu) \right] =  g(\mu)^\delta \ , \quad g(\mu) \ll 1 \ , \label{weakN1} \\
&& \Gamma\left[ N(\mu), g(\mu) \right] = g(\mu)^{-\delta} \ , \quad g(\mu) \gg 1   \ , \quad \delta > 0 \ . \label{weakN2}
\end{eqnarray}
We will, however, not explore this case in this article.

To proceed further, we need to relate the time co-ordinate with the energy-scale at which measurement is being carried out. On dimensional ground, $t = \xi/\mu$, where $\xi $ is an undetermined constant. However, this parameter is physical. When $\xi \sim \O(1)$, one is making dynamical observations at the scale set by the inverse energy and hence there is no decoupling between time evolution and coarse graining. On the other hand, setting $\xi \gg 1 $ decouples the two. It is in this sense, given a state at a particular energy-scale, one can measure long-time behaviour staying at the same energy scale. The resulting Callan-Symanzik equation takes the form:
\begin{eqnarray}
\partial_\mu \lambda_L = -\beta \left(  \frac{\alpha}{\xi g} + \frac{\partial_g \lambda_L}{\mu} \right) + \frac{\lambda_L}{\mu} - \lambda_L \frac{\partial_\mu \xi}{\xi}+ \frac{\partial_\mu N}{N} \left( \frac{\mu}{\xi} - N \partial_N \lambda_L\right)  \ . \label{CSLya}
\end{eqnarray}
Let us consider some generic possibilities. Note that $\partial_\mu N \le 0$, along an RG-flow.

Consider a vanishing beta function. This case has a very simple exact solution: $\lambda_L = \mu \log N$, with $\xi=1$. Note that this is a monotonically increasing function of the energy-scale. Also, physically, this corresponds to a highly evolving state and the corresponding result cannot be used for a thermal state. It is, nonetheless, interesting that a highly evolving state can readily exhibit an exponentially growth in time. On the other hand, in the $\xi \gg 1$ limit, we can choose $\xi = N$. Upon further using the condition that $\partial_N \lambda_L \to 0$ compared to all variation in the system, the Callan-Symanzik equation can be solved:
\begin{eqnarray}
\lambda_L = c_1 \frac{\mu}{N} + \mu \frac{\log N}{N} \ . \label{Lyaweak}
\end{eqnarray}
In the large $N$ limit, this yields a small Lyapunov exponent, much like what is observed at weak coupling in \cite{Stanford:2015owe, Steinberg:2019uqb} for the thermal state. It is conceivable that the $\log N$ factor in the numerator is present at the fixed point, which is absent in \cite{Stanford:2015owe, Steinberg:2019uqb} which hold an a non-vanishing beta. Also note that, in this case $\partial_\mu \lambda_L > 0$ in the large $N$ limit.

To make the comparison with known results for a thermal state more precise, let us consider the following cases: First, assume that the beta function vanishes and $\partial_\mu N(\mu)$ is small compared to all variations in the system and we can safely ignore the $N'(\mu)$-terms from the Callan-Symanzik equation. The resulting equation can simply be solved to obtain:
\begin{eqnarray}
\lambda_L (\mu) = \mu c_1 \ , \label{Lyastrong}
\end{eqnarray}
where $c_1$ is an integration constant. Given a thermal state for a large $N$ theory with a Holographic dual\cite{Maldacena:2015waa}, we can identify $\mu = T$ and $c_1 = 2\pi$, where $T$ is the corresponding temperature. In this case, along an RG-flow $\partial_\mu \lambda_L > 0$, assuming positivity of the Lyapunov exponent itself. Thus, the Lyapunov decreases towards the IR.

Let us make a couple of more observations from (\ref{CSLya}). First, note that the solution in (\ref{Lyaweak}) is always positive in the large $N$ limit, assuming that integrations constants do not scale with such large numbers. The philosophical basis of this assumption is rooted in the premise that the integration constant is unaware of the degrees of freedom in the system. However, the solution in (\ref{Lyastrong}) is not necessarily positive and everything depends on the integration constant. In the former case, the correlator shows ergodic behaviour by virtue of being a large $N$ system, while in the latter this is not the case. Thus, $\partial_N \lambda_L \to 0$ can be seen as a necessary condition for ergodicity in correlators, near a fixed point.

Suppose now, we want to consider a complex solution of (\ref{CSLya}). Physically, this implies an exponential dependence on time, as well as an oscillatory part. The equation (\ref{CSLya}), at vanishing beta function, can then be re-written in terms of a real and a purely imaginary equation. These take the form:
\begin{eqnarray}
&& \partial_\mu \lambda_L^{R} =  \frac{\lambda_L^R}{\mu} - \lambda_L^R \frac{\partial_\mu \xi}{\xi}+ \frac{\partial_\mu N}{N} \left( \frac{\mu}{\xi} - N \partial_N \lambda_L^R \right)  \ . \label{CSLyaR} \\
&&  \partial_\mu \lambda_L^{I} =  \frac{\lambda_L^I}{\mu} - \lambda_L^I \frac{\partial_\mu \xi}{\xi} - \partial_\mu N \partial_N \lambda_L^I \ , \label{CSLyaI}
\end{eqnarray}
where $\lambda_L = \lambda_L^R + i \lambda_L^I$. Note that, if we now impose $\partial_N\lambda_L \to 0$, the above system always has a solution with $\lambda_L^I =0$ and $\lambda_L^R \not = 0$. This corresponds to an exponential behaviour in time, and ergodic for $\lambda_L^R > 0$, as we have also argued above. On the other hand, suppose the correlator is purely oscillatory, then $\lambda_L^R =0$. This, however, does not necessarily impose $\partial_\mu \lambda_L^R = 0$ or $\partial_N \lambda_L^R = 0$. From the equation (\ref{CSLyaR}), we observe that a consistent solution of $\lambda_L^R$ needs to have non-vanishing derivatives and therefore will become non-vanishing away from the fixed point. This predicts that, even if we start with a fixed point where OTOCs are oscillatory, slightly away from this fixed point they will develop real exponential dependence.

The reverse is not true. Setting $\lambda_L^I =0$ along with $\partial_\mu \lambda_L^I =0$ and $\partial_N\lambda_L^I =0 $ is certainly consistent with (\ref{CSLyaI}). Thus, if we begin with an ergodic fixed point, the physics slightly away from it also remains ergodic.

Let us now perform a near fixed point analyses. For this, we can generically assume distinct expansions for the coupling constant and the degrees of freedom, in terms of the energy. For example:
\begin{eqnarray}
g(\mu) = g_0 + \sum_{i=1} \epsilon^i \mu^i g_i \ , \quad N(\mu) = N_0 +  \sum_{i=1} \epsilon^i \mu^i N_i  \ , \quad \beta = \mu \partial_\mu g(\mu) \ , 
\end{eqnarray}
where $\epsilon $ is the dimensionless expansion parameter. Consider now the case when $N_i = 0$, but $g_1 \not = 0$. At the leading order, one obtains:
\begin{eqnarray}
\lambda_L = c_1 \mu + \epsilon \left( c_2 \mu - \frac{g_1}{g_0} \mu^2 \frac{\alpha}{N_0} \right) \ , 
\end{eqnarray}
where $c_1$ and $c_2$ are two integration constants. Given the solution above, $\lambda_L \sim \mu$ follows simply from scaling and thus we can interpret $(c_1 + \epsilon c_2)$ as the corresponding proportionality constant. Assuming $\alpha$ is positive,\footnote{This is reasonable in a perturbative regime, and $\alpha<0$ clearly cannot be captured within the perturbative framework, without a possible resummation.} we see that the change in $\lambda_L$ is tied to the sign of the beta function, since $\beta \sim g_1$. For a positive beta-function, the Lyapunov decreases while for a negative beta function it increases.

We will now consider systems which saturate the chaos bound. As we have seen, the Lyapunov is a decreasing function towards the IR. We will instead consider how the scrambling time behaves as one tunes from the UV to the IR.

\section{Holographic QFT: Examples}

Now we will consider explicit holographic QFTs, including CFTs, in which certain degrees of freedom scramble faster in the IR while certain others scramble slower in the IR. These examples consist of SU$(N_c)$ strongly coupled gauge theories, with $N_q$ matter in the fundamental representation. The adjoint matter obeys a monotonically decreasing scrambling time, while the fundamental matter does not. The reason is simple: while the adjoint matter scrambling is independent of the gauge coupling, for the fundamental matter, it is not so.

\subsection{D$p$-brane Geometries}

We will follow the notation of \cite{Itzhaki:1998dd}. Let us start with $10$-dimensional closed-string geometry, sourced by a stack of D$p$-branes, in the string frame:
\begin{eqnarray}
ds^2 & = & \left( \frac{u}{L}\right)^{\frac{7-p}{2}} \left[ - f(u) dt^2 + dx_p^2 \right] +   \left( \frac{L}{u}\right)^{\frac{7-p}{2}} \frac{du^2}{f(u)} + \left( \frac{L}{u}\right)^{\frac{7-p}{2}} u^2 d\Omega_{8-p}^2 \ , \label{csmet1} \\
e^\phi & = &\left( \frac{u}{L}\right)^{\frac{(p-3)(7-p)}{4}} \ , \quad f(u) = 1 - \left( \frac{u_H}{u}\right)^{7-p}\label{csmet2} \ , \\
F_{8-p} & = & (7-p) L^{7-p} \omega_{8-p} \ . \label{csmet3}
\end{eqnarray}
Here, $\omega_n$ is the volume form of the $n$-sphere, $L$ is the overall curvature scale of the geometry, $u_H$ denotes the location of the event horizon. The curvature scale, $L$ is fixed by the Dirac quantization relation:
\begin{eqnarray}
\int F_{8 - p} = 2 \kappa^2 T_{Dp} N \ , 
\end{eqnarray}
where 
\begin{eqnarray}
\frac{1}{2\kappa^2} = \frac{2\pi} {g_s^2 \left( 2 \pi \ell_s \right)^8} \ , \quad T_{Dp} = \frac{1}{g_s \ell_s \left( 2\pi \ell_s\right)^p} \ , \label{rel1}
\end{eqnarray}
where $g_s$ is the string coupling constant and $\ell_s$ is the string length. Here $(2\kappa^2)$ denotes the ten-dimensional Newton's constant and $T_{Dp}$ is the corresponding D-brane tension. This relation fixes the curvature scale with string coupling and string length:
\begin{eqnarray}
L^{7-p } = \frac{\left( 2 \pi \ell_s \right)^{7-p}}{(7-p) V_{8-p}} g_s N \ . \label{rel2}
\end{eqnarray}

The putative dual field theory for these geometries are given by $(p+1)$-dimensional maximally supersymmetric SU$(N)$ gauge theories, with purely adjoint degrees of freedom, with a gauge coupling:
\begin{eqnarray}
g_{\rm YM}^2 = 2\pi g_s \left(2 \pi \ell_s \right) ^{p-3} \ . \label{gym}
\end{eqnarray}
The corresponding 't Hooft coupling is defined as: $\lambda= g_{\rm YM}^2 N$. To introduce further dynamical degrees of freedom, one can introduce an open string sector. This can be done by introducing an open string itself, or by an appropriate D-brane, which corresponds to introducing a set of fundamental matter sector. Note further that, in (\ref{csmet1})-(\ref{csmet2}), the co-ordinates carry dimension of length. 

Now, the geometric description is valid, provided $g_s e^\phi \ll 1$, to avoid string coupling corrections and $\ell_s^2 \R \ll 1$, to suppress higher curvature corrections. Here $\R$ is the curvature of the ten-dimensional geometry. Using the geometric data in (\ref{csmet1})-(\ref{csmet3}), these conditions evaluate to be: 
\begin{eqnarray}
&& g_s e^\phi \sim \frac{1}{N} \left( \lambda \, \E_{\rm UV}^{p-3}\right)^{(7-p)/2} \ , \\
&& \ell_s^2 \R \sim \left(\lambda \, \E_{\rm UV}^{p-3} \right)^{-1} \ , \quad \E_{\rm UV} = \frac{u}{\ell_s^2}\ , 
\end{eqnarray}
where $\E_{\rm UV}$ is an energy-scale, relevant for the UV-physics.\footnote{Note that, this is not a gauge-invariant way of assigning an energy-scale, only one scheme of defining an energy-scale. However, for $p =3$, which is conformal, this is indeed as good as any other energy-scale definition. We will not dwell upon the gauge-invariant definition of energy for $p\not =3$, since our final physical results will not be sensitive to this.} The validity of the geometric description is hence given by
\begin{eqnarray}
N^{4/(7-p)} \gg \lambda \, \E_{\rm UV}^{p-3} \gg 1 \ .
\end{eqnarray}
A particularly interesting case is $p=3$, in which the above hierarchy becomes: $N^2 \gg \lambda \gg 1$.

The scrambling time can be estimated\footnote{See {\it e.g.}~\cite{Shenker:2013pqa, Shenker:2013yza, Shenker:2014cwa}.} by computing the back-reaction of a probe particle, with a fixed energy $E$ at the UV, at time $t_0$.  For late times, $t_0 > 1/T$ where $T$ is the temperature of the black hole, the energy of the particle is blueshifted by an exponential factor: $E e^{2\pi t_0 T}$. The presence of the event horizon washes away any other details of the system. Now, one simply evaluates the gravitational back-reaction of the blueshifted mode on the original spacetime. The stress-tensor will be weighted by a factor of the ten-dimensional Newton's constant: $\kappa^2$. Upon calculating the on-shell Euclidean supergravity action, with this back-reaction, one obtains the relevant Euclidean correlator and subsequently analytical continuation yields the corresponding OTOC. The schematic result is simply:
\begin{eqnarray}
f(t) = 1 - \Gamma \kappa^2 e^{2\pi T t} \ , \quad t_{\rm d} \ll t \ll t_{\rm scr} \ ,
\end{eqnarray}
where $t_{\rm d} \sim 1/T$ is the dissipation time-scale and $t_{\rm scr}$ is the scrambling time. Here, $\Gamma$ knows about the details of the geometry, including the large parameters of the state. The parametric scaling form of $\Gamma \kappa^2$ can be obtained by evaluating:
\begin{eqnarray}
\frac{1}{2\kappa^2} \int d^{10}x \sqrt{-G} e^{-2\phi} \R \sim \frac{1}{g_s^2 \ell_s^8}  \sim N^2 \lambda ^{\frac{3- p}{ p-5}} \E_{\rm UV}^{\frac{3-p}{p-5}} = \Gamma\kappa^2 \ . \label{gammakappa}
\end{eqnarray}
The scrambling time is defined as: $f(t_{\rm scr}) = 0$, which yields:
\begin{eqnarray}
t_{\rm scr} = \frac{1}{2\pi T} \log \left( N^2 \lambda ^{\frac{3- p}{ p-5}} \E_{\rm UV}^{\frac{3-p}{p-5}} \right) \ . \label{DpscrUV}
\end{eqnarray}
In evaluating (\ref{gammakappa}), we have used that the overall scaling of $ \sqrt{-G} e^{-2\phi} \R $, on-shell, is given by an appropriate power of $L$. 

The same answer is obtained following \cite{Shenker:2013pqa}. In this one begins with a thermofield double description of the geometry, which is certainly available for the background in (\ref{csmet1})-(\ref{csmet3}). Now, one considers injecting a small amount of energy, denoted by $E$, to the left QFT, at a time $t = - t_{\rm in}$, where $t_{\rm in} >0$. At a later time $t=0$, this energy will blueshift to $E e^{2\pi T t_{\rm in}}$, where $T$ is the temperature of the black hole phase of (\ref{csmet1})-(\ref{csmet3}). This energy should be measured, naturally, in the units of the black hole mass, $M$ itself and therefore the relevant quantity controlling the back-reaction of the injected energy at a later time is given by $(E/M) e^{2\pi T t_{\rm in}}$. Now, one simply assigns the minimal natural energy for the black hole, proportional to the Hawking temperature: $E \sim T$, and therefore obtains the corresponding scrambling time by setting: $(E/M) e^{2\pi T t_{\rm in}} \sim \O(1)$. This yields: $t_{\rm scr} \sim \frac{1}{2\pi T} \log S$, where $S$ is the entropy of the black hole. For our purposes, we will safely replace the $\sim$ with an exact equality, since we will not care about order one constants.

It is straightforward to calculate the entropy of the black hole geometry in (\ref{csmet1})-(\ref{csmet3}), by evaluating the horizon area in the Einstein-frame. The Einstein-frame metric is given by $G_{\mu\nu}^E = e^{-\phi/2} G_{\mu\nu}^s$, where $G_{\mu\nu}^s$ is given in (\ref{csmet1})-(\ref{csmet3}). The entropy is given by
\begin{eqnarray}
S = N^2 \lambda^{\frac{3-p}{p-5}} T^{\frac{9-p}{5-p}} \Z_1 \ , 
\end{eqnarray}
where $\Z_1$ is an order one dimensionful constant. Clearly, $\frac{1}{2\pi T} \log S$ matches with (\ref{DpscrUV}). The basic observation to note here is that the scrambling time depends on the number of degrees of freedom as well as the 't Hooft coupling, except at $p=3$, when the coupling disappears.\footnote{Note that, we define the number of degrees of freedom as simply the coupling-independent, scale-independent factor that extensive thermodynamic functions scale with. Thus, here, the number of degrees of freedom is simply $N_c^2$.} This happens simply because for $p=3$ the dynamics is conformal and no scales appear in the resulting scrambling time. This observation is highly suggestive that for this case, since $\lambda$ provides us with no scale, the scrambling behaviour of $\log N$ is robust against corrections in $1/\lambda$, perturbatively. For any other $p$, however, this is not true.

\subsection{Localized Probes}

For such large $N$ SU$(N)$ gauge theories, a generic class of measurements can be carried out on an appropriate probe field, {\it e.g.}~like a probe quark in the bath of gluons. Such degrees of freedom are realized by introducing open string degrees of freedom in the bulk geometry. This can be done by considering various D-brane probes in the given supergravity geometry, see {\it e.g.}~\cite{Mateos:2006nu, Albash:2006ew, Karch:2006bv, Albash:2007bq, Mateos:2007vn, Karch:2007pd, Johnson:2008vna, Alam:2012fw, Kundu:2013eba} for thermal physics on this probe sector but the simplest is to consider a long open string as a probe itself. The dynamics is governed by an effective thermal physics on the worldsheet or worldvolume of such degrees of freedom\cite{Kim:2011qh, Kundu:2015qda, Kundu:2018sof, Kundu:2019ull}.

A fundamental string, in the probe limit, can be introduced such that the worldsheet coordinates are stretched along $\{t, u\}$-submanifold of the geometry, and the embedding function is described by $x_p(t, u)$, in general. The dynamics is described by
\begin{eqnarray}
&& S_{\rm NG} = - \frac{1}{2\pi \ell_s^2} \int dt du \sqrt{- {\rm det} \gamma} \ , \label {act1}\\
&& \gamma_{ab} = G_{\mu\nu} \left( \partial_a X^\mu \right)  \left( \partial_b X^\nu\right)  \ ,
\end{eqnarray}
where $G_{\mu\nu}$ is the background metric and $X^\mu$ are the embedding function. Given a spacetime, one can find out certain specific classical embedding, by solving for $x_p(t, u)$ from the Nambu-Goto equations of motion. Now, action (\ref{act1}) defines a classical theory which we may consider quantizing semi-classically. This can be implemented by considering the quadratic fluctuations around the classical profile as the quantum field theory. While $x_p(t,u)$ can be non-trivial classical functions, for the geometric background in (\ref{csmet1}), it is easy to check that $x_p(t,u)={\rm constant}$ is always a solution, simply warranted by homogeneity of the background. Now, the schematic form of the quantum field theory takes the form:
\begin{eqnarray}
S_{\rm quad} = - \frac{1}{2\pi \ell_s^2} \int dt du \sqrt{- {\rm det}\gamma} \left( \gamma^{ab} G_{xx} \left( \partial_a \delta X \right)  \left( \partial_b \delta X\right) + M_x^2 \delta X \delta X \right) \ , \label{quadNG}
\end{eqnarray}
where $M_X^2$ is the effective mass around this classical profile. The action above is essentially a Gaussian QFT, which we can now quantize and obtain the free propagator. To consider $4$-point, or any higher point, correlation function, one essentially  expands the action to the desired order in $\delta X$: 
\begin{eqnarray}
S_{\rm NG-int} = S_{\rm quad} + S_{\rm int} \left[ \left(\delta X\right)^3 , \left(\delta X\right)^4 , \ldots \right]  \ , 
\end{eqnarray}
and using the path-integral formalism, simply takes $n^{\rm th}$ derivative of the path integral with respect to the sources, and sets the sources to zero:
\begin{eqnarray}
&& \left \langle \O_1 \O_2 \ldots \O_n \right \rangle  = \left. \frac{1}{Z_{\rm quad}} \frac{\delta}{i \delta (\delta X_1)} \ldots \frac{\delta}{i \delta (\delta X_n)} Z_{\rm int} \right|_{\delta X_{\rm quad}} \ ,\\
&& Z_{\rm quad} = \int \left[\D \delta X\right] e^{i S_{\rm quad}} \ , \\
&& Z_{\rm int} = \int \left[\D \delta X\right] e^{i S_{\rm NG-int}} \ . 
\end{eqnarray}

Now, the quadratic piece of the quantized action can be canonically normalized by absorbing one factor of $\ell_s^{-1}$ in to $\delta X$, which makes the action (\ref{quadNG}) independent of $\ell_s$. As a result, the quartic interaction will now have an overall factor of $\ell_s^2$ as the coupling. This now becomes a small coupling expansion, in positive powers of $\ell_s$. Subsequently, the $4$-point function, appropriately normalized, picks up a coefficient $\ell_s^2$ whose inverse is the effective tension of the probe and it determines the scrambling time, see {\it e.g.}~\cite{Murata:2017rbp, deBoer:2017xdk, Banerjee:2018kwy, Banerjee:2018twd, Banerjee:2019vff}.

The behaviour of the $4$-point OTOC, in the thermal state is given by
\begin{eqnarray}
\O(t) = 1  - \left( {\rm Tension}^{-1} \right) \left( {\rm Dynamical}\right)  \ , 
\end{eqnarray}
where $\left( {\rm Tension} \right)$ denotes the effective tension of the probe and $\left( {\rm Dynamical}\right)$ denotes the time evolving behaviour. Here, the effective inverse tension is simply determined by $2\pi \ell_s^2$ and, for a thermal state, the dynamical contribution is an exponential growth. Using the relations in (\ref{rel1}) and (\ref{rel2}), the scrambling time is thus estimated as:
\begin{eqnarray}
\O(t_{\rm scr}) \sim \O(1) \quad \implies \quad t_{\rm scr} = \frac{\alpha}{\lambda_{\rm L}} \log \left( \lambda \E^{p-3} \right) \ , \label{UVscr}
\end{eqnarray}
where $\lambda_{\rm L}$ is the Lyapunov exponent, and $\alpha$ is a positive constant, depending on various details of the geometry. Furthermore, $\E$ denotes a typical energy scale at which we are probing the system. Such a factor arises purely on dimensional grounds, since the 't Hooft coupling is dimensionful, except for $p=3$. The basic result here is that, for such strongly coupled large $N$ gauge theories, fundamental degrees of freedom scramble information at a logarithmic rate with respect to the 't Hooft coupling, independent of the otherwise degrees of freedom of the system which scales as $N^2$. Finally, note that, the meaningful quantity here is $\left( t_{\rm scr} \lambda_{\rm L} \right)/\alpha$ which has a good logarithmic definition. This is what we will consider in this article.

At this level, the observation in (\ref{UVscr}) is suggestive of an energy-scale dependence of the corresponding scrambling time. Any question related to the thermal physics can further be characterized in terms of the energy scale at which thermalization is taking place. This can happen in QFT quite naturally, since a given ``UV QFT" can thermalize itself, or it can first flow non-trivially to an ``IR QFT" and then thermalize. While at conformal fixed points, the corresponding scrambling time is only sensitive to the number of degrees of freedom, away from it, this depends on the associated energy-scale. As a result, one freely obtains an energy-scale $\E^{p-3}\sim \lambda^{-1}$, at which the scrambling time is order one. The last observation is highly suggestive of a ``loss of semi-classical" limit, which is simply the fact that one needs to now include the higher curvature corrections\footnote{Equivalently, include higher spin degrees of freedom in the string spectrum.} in the geometric description.\footnote{Note, however, that all gravitational fluctuations will couple with the effective Newton's constant of the geometry, therefore will only be sensitive to the total number of degrees of freedom. For them, higher curvature corrections in the geometry will not alter the scrambling time. Said differently, it will not be surprising if the higher derivative theories of gravity still allow the Einstein-gravity shock-wave geometry as a solution.} See {\it e.g.}~\cite{Alishahiha:2016cjk} for similar physics in candidate higher derivative gravities. 

\subsection{Smeared Probes}

Let us now carry out a similar analysis, where we introduce a small number of probes and smear them along the transverse directions. The reason behind this is, momentarily, we will consider geometries that arise from the back-reaction of such smeared string sources, on the D$p$-brane geometries. The action of the smeared probe sector is given by
\begin{eqnarray}
S_{\rm smear} = - \frac{1}{2\pi \ell_s^2} \int dt du \sqrt { - {\rm det} \gamma} \int_{\Sigma_8} d\sigma_8 \ ,
\end{eqnarray}
where $d\sigma_8$ is the smearing form. On dimensional ground, the smearing form should be proportional to the volume form, up to an overall dimensionless number. The dimensionless number plays no role in our analyses, assuming that this number does not scale with $N$ or $\lambda$ or any such physical large parameters in the system.\footnote{In fact, to ensure the probe limit, this number should be much smaller compared to $N$ or $\lambda$.} Therefore, let us take $d\sigma_8$ to be the volume form. 

Now, the effective tension of the probe sector is given by
\begin{eqnarray}
T_{\rm eff} \sim \frac{1}{\ell_s^2} L^{\frac{7-p}{2}(4-p)} \ ,
\end{eqnarray}
where we see that the string tension $\ell_s^{-2}$ is accompanied by an overall curvature scale, raised to an appropriate power. We are free to set this curvature scale to any order one numerical value, which sets the length scale in the gravitational description. Thus, the effective tension only receives an order one correction due to smearing. Using the definition of the gauge coupling in (\ref{gym}), (\ref{rel2}) and the definition of the 't Hooft coupling, the estimated scrambling time is again given by
\begin{eqnarray}
t_{\rm scr} \sim \frac{\alpha_{\rm smear}}{\lambda_L} \log \left( \lambda \E^{p-3}\right) \ , \label{scrsmear}
\end{eqnarray}
where $\alpha_{\rm smear}$ is a numerical constant, and $\E$ is the typical energy-scale at which we are probing the system. Note that the scrambling times in (\ref{UVscr}) and (\ref{scrsmear}) only differ by the order one numerical constant, captured by $\alpha$ and $\alpha_{\rm smear}$. Stated differently, at the UV, smeared degrees of freedom and localized degrees of freedom scramble in a qualitatively similar manner. We will momentarily see that this is not the case at the IR.

\section{An Infra-red Geometry}

In the previous section we have discussed how scrambling time behaves for the bulk open string degrees of freedom, or the boundary fundamental matter sector. The probe limit remains valid only till an IR energy-scale, beyond which the geometry is significantly different, and in fact, non-perturbative in back-reaction. Given a UV D$p$-brane geometry in (\ref{csmet1})-(\ref{csmet3}), the back-reacted IR geometry, in the string frame, is given by
\begin{eqnarray}
&& ds^2 = e^{\phi/2} \left[ e^{- 2 \frac{8-p}{p} \eta} ds_{p+2}^2 + e^{2\eta} L^2 d\Omega_{8-p}^2 \right] \ , \\
&& ds_{p+2}^2 = g_{tt} dt^2 + g_{rr} dr^2 + g_{xx} dx_p^2 \ , \\
&& e^\phi = \beta_\phi Q^{\frac{p-7}{2}} \left( \frac{r}{L}\right)^{\frac{p(p-7)}{2(p-4)}} \ , \quad e^{2\eta} = \beta_\eta Q^{\frac{3-p}{4}} \left( \frac{r}{L} \right)^{\frac{p(3-p)}{4(p-4)}} \ ,
\end{eqnarray}
where $\beta_{\eta,\phi}$ are numerical constants, which we need not keep track of. The explicit form of $ds_{p+2}^2$ is given below:
\begin{eqnarray}
&& ds_{p+2}^2 = \left( \frac{r}{L}\right)^{-\frac{2\theta}{p}} \left[ - \left(\frac{r}{L} \right)^{2z} f(r) dt^2 +  \left(\frac{r}{L} \right)^{2} dx_p^2 + \beta_\ell Q^{\frac{2(3-p)}{p}} \left( \frac{L}{r}\right)^2 \frac{dr^2}{f(r)} \right] \ , \label{backmet1} \\
&& f(r) = 1 - \left( \frac{r_H}{r}\right)^{p-\theta+z} \ , \\
&& Q \sim \ell_s^{4 \frac{6- p}{7-p}} \lambda^{\frac{8-p}{7-p}} \left( \frac{N_q}{N^2} \right)  \ , \quad z = \frac{16-3p}{4-p} \ , \quad \theta = \frac{p(3-p)}{4-p} \ . \label{backmet2} 
\end{eqnarray}
where $\beta_\ell$ is a purely numerical constant and $N_q$ is the back-reacting string density. We have also ignored the overall numerical constant relating $Q$ and $\{\ell_s, \lambda, N_q, N \}$. There is a special case: $p=4$, for which the background geometry takes the following form:
\begin{eqnarray}
&& ds_{6}^2 = \left( \frac{r}{L}\right) ^{1/2} \left[ - \left( \frac{r}{L}\right) ^{2} dt^2 + dx_4^2 + \beta_\ell^2 Q^{-1/2} \left( \frac{L}{r}\right) ^{2} dr^2 \right]  \ , \label{backmet3} \\
&& e^\phi = \beta_\phi Q^{-3/2} \left( \frac{r}{L}\right) ^{3/2} \ , \quad e^{2\eta} = \beta_\eta Q^{-1/4} \left( \frac{r}{L}\right) ^{1/4} \ , \label{backmet4} 
\end{eqnarray}

Before discussing probe strings, let us estimate the scrambling time for gravitational degrees of freedom in these geometries. As explained before, this amounts to estimating the entropy of the corresponding black hole solution, which using (\ref{backmet1})-(\ref{backmet4}) is found to be:
\begin{eqnarray}
t_{\rm scr} \sim \frac{1}{\lambda_{\rm L}} \log \left[ N^2 \left(\frac{N_q}{N^2}\right)^{2(6-p)/(16-3p)} \lambda^{\frac{p}{16-3p}}\right] \ , \label{HVscr}
\end{eqnarray}
Comparing (\ref{DpscrUV}) and (\ref{HVscr}), with $N_q / N^2 \sim \O(1)$, one obtains: $\lambda_L^{\rm IR} t_{\rm scr}^{\rm IR} > \lambda_L^{\rm UV} t_{\rm scr}^{\rm UV}$. This inequality seems to generically persist in the models we consider here. Note further that, the IR geometry is sourced by certain UV degrees of freedom that we have explicitly introduce. The number of degrees of freedom that are relevant for estimating the scrambling time increase from the UV to the IR, since the only in the IR the additional degrees of freedom explicitly introduced at the UV become relevant.

\subsection{Scrambling on a Localized Probe}

First, the IR background data yields the following: 
\begin{eqnarray}
g_s e^\phi \sim \frac{N_c^{6-p}}{N_q^{(7-p)/2}} r^{\frac{p(7-p)}{2(4-p)}} \lambda^{\frac{(7-p)(p-3)}{2(5-p)}} \ , \quad \ell_s^2 R_{10} \sim - \frac{N_q}{N_c^2} r^{- \frac{p}{4-p}} \lambda^{\frac{3-p}{5-p}} \ .
\end{eqnarray}
Validity of type II supergravity requires $g_s e^\phi \ll 1$ and $\ell_s^2 R_{10} \ll 1$. These two translate into:
\begin{eqnarray}
r^{\frac{p(7-p)}{2(4-p)}} \ll  \left( \frac{N_q}{N_c^2} \right)^{\frac{7-p}{2}} N_c \lambda^{\frac{(7-p)(3-p)}{2(5-p)}}   \ , \quad r^{\frac{p(7-p)}{2(4-p)}} \gg  \left( \frac{N_q}{N_c^2} \right)^{\frac{7-p}{2}}  \lambda^{\frac{(7-p)(3-p)}{2(5-p)}} \ . \label{sugravalidgen}
\end{eqnarray}
We will need these constraints to proceed further. Equivalently, the above window defines the energy-band within which our supergravity description holds. To estimate this energy-scale, we need to relate the IR radial coordinate, with the energy-scale in the UV QFT. This can be done by demanding that the string density remains the same along the RG-flow from the UV to the IR and one obtains:
\begin{eqnarray}
&& \log\left( \frac{r}{L} \right) = \frac{(9-p)(4-p)}{2p} \log \left( \frac{\E \ell_s^2}{L}\right)  \ , \quad p \not = 4 \ , \\
&& \log\left( \frac{r}{L} \right) = \frac{5}{2} \log \left( \frac{\E \ell_s^2}{L}\right)  \ , \quad p = 4 \ , 
\end{eqnarray}
where $\E= u/ \ell_s^2$ defines a convenient energy-scale for the UV QFT. Now, the allowed energy band is given by
\begin{eqnarray}
\lambda^{\frac{3}{9-p}} \left( \frac{N_q}{N_c^2} \right)^{\frac{2}{9-p}} \ll \E_{\rm IR} \ll \lambda^{\frac{3}{9-p}} \left( \frac{N_q}{N_c^2} \right)^{\frac{2}{9-p}}  N_c^{\frac{4}{(7-p)(9-p)}} \ .
\end{eqnarray}
For example, for $p=3$, the above range is given by
\begin{eqnarray}
\left( \frac{N_q}{N_c^2}\right)^{1/3} \sqrt{\lambda} \ll \E_{\rm IR} \ll N_q^{1/3} \sqrt{\frac{\lambda}{N_c}} \ .
\end{eqnarray}

Let us consider now a probe string in the background geometry of (\ref{backmet1})-(\ref{backmet4}). To describe a classical profile, for a string which is stretched along the radial direction of the geometry, we can pick a function $x_p(r)$ that extremizes the Nambu-Goto equations of motion. For the geometries described above, it is easy to check that $x_p(r) = {\rm const}$ is a classical solution. Now, in the semi-classical regime, we consider fluctuations around the classical profile. We will canonically normalize the the quadratic kinetic piece, and subsequently, evaluate the quartic piece in the classical background, setting further the quadratic piece on-shell. If the worldsheet has an event horizon, which is the case when the background has an event horizon, it is thus expected that the quartic piece will yield a four-point correlation function. By performing an appropriate analytic continuation, we can further calculate the four-point out-of-time-order correlator that will exhibit an exponential growth in real time.

As outlined in the previous section, the corresponding scrambling time can be estimated simply by reading off the inverse effective tension of the probe degree of freedom. For the geometries in (\ref{backmet1})-(\ref{backmet4}), this yields:
\begin{eqnarray}
t_{\rm scr} & = &  \frac{1}{\lambda_L^{\rm IR}}   \log \left[ \left( \frac{N_c^2}{N_q} \right)^{\frac{3}{p}} \lambda^{- \frac{4(3-p)}{p(5-p)}}  \E^{\frac{36 - 9 p + p^2}{p(5-p)}} \right] \ .
\end{eqnarray}
One can easily compare this with the scrambling time at the UV, in (\ref{UVscr}). Note that, in the UV, the factor of the 't Hooft coupling always guarantees the scrambling time to parametrically separate from earlier time-scales, {\it e.g.}~the dissipation time-scale. A similar statement is true in the IR as well, provided one takes into account the constraints in (\ref{sugravalidgen}). Interestingly, what in the UV was solely determined by the 't Hooft coupling, in the IR, involves a non-trivial combination of all large numbers in the problem. The only exception is at $p=3$, in which the IR-scrambling is independent of the 't Hooft coupling.

\subsection{Scrambling on a Smeared Probe}

Suppose now, we consider smearing the additional probe strings uniformly, as is done to obtain the back-reacted geometry at the first place. This will require us to consider the following action for the probe:
\begin{eqnarray}
S_{\rm smear} = - \frac{1}{2\pi \ell_s^2} \int dt dr \sqrt{\- g_{tt} g_{rr}} \int_{\Sigma_8} d\sigma_8 \ , 
\end{eqnarray}
where $\Sigma_8$ is the transverse manifold to the string worldsheet and $d\sigma_8$ is the volume form on it. The smearing clearly comes along with an additional volume factor, and therefore the effective tension of this degree of freedom will change. Based on the geometry in (\ref{backmet1})-(\ref{backmet2}), one can easily evaluate the scrambling time as:
\begin{eqnarray}
t_{\rm scr} = \frac{1}{\lambda_L^{\rm IR}}  \log \left[ \left( \frac{N_c^2}{N_q} \right)^{7 - p + \frac{3}{p}} \lambda^{- \frac{12+ (8-p)(3-p)p}{p(5-p)}}  \E^{\#} \right] \ ,
\end{eqnarray}
where $\#$ is some number that ensures the argument inside the logarithm is dimensionless. One can again check that, a large scrambling time is guaranteed, upon using the conditions in (\ref{sugravalidgen}). In fact, this is always ensured by the low curvature condition. Thus, one obtains the following generic relation:
\begin{eqnarray}
\alpha_1 \lambda_L^{\rm IR}t_{\rm IR}^{\rm loc} = \alpha_2  \lambda_L^{\rm IR} t_{\rm IR}^{\rm smr} - \alpha_3  \lambda_L^{\rm UV} t_{\rm UV}^{\rm loc/smr} \ ,
\end{eqnarray}
where $\alpha_{1,2,3}$ are all positive numbers and $t_{\rm IR}$ and $t_{\rm UV}$ denote the IR and the UV scrambling times, respectively.

As a particular example, let us consider $p=3$. This yields:
\begin{eqnarray}
\lambda_L^{\rm IR} t_{\rm IR}^{\rm smr} = 5 \lambda_L^{\rm IR} t_{\rm IR}^{\rm loc} - 4 \lambda_L^{\rm UV} t_{\rm UV}^{\rm loc}  \ . \label{scrrel}
\end{eqnarray}
Now, consider the time-scale differences between UV localized probes and IR localized probes:
\begin{eqnarray}
\lambda_L^{\rm IR} t_{\rm IR}^{\rm loc} - \lambda_L^{\rm UV} t_{\rm UV}^{\rm loc} = \log \left( \frac{N_c^2}{N_q} \E_{\rm IR}^3 \frac{1}{\lambda} \right) \ ,
\end{eqnarray}
where $\E_{\rm IR}$ is the IR energy-scale. It is easy to check that, given the UV-geometry, the backreacted IR region opens up precisely at the point where the string back-reaction becomes order one. This provides one with the cross-over energy-scales below $\E_{\rm cross}$:
\begin{eqnarray}
\E_{\rm cross}^3  = \lambda^2 \frac{N_c^2}{N_q} \ .
\end{eqnarray}
Now, for any $\E_{\rm IR} \le \E_{\rm cross}$, one obtains: $\lambda_L^{\rm IR} t_{\rm IR}^{\rm loc} > \lambda_L^{\rm UV} t_{\rm UV}^{\rm loc}$, and therefore scrambling time increases in the IR.

In the same spirit, let us consider the case of generic $p$. One obtains:
\begin{eqnarray}
\lambda_L^{\rm IR} t_{\rm IR}^{\rm loc}  - \lambda_L^{\rm UV} t_{\rm UV}^{\rm loc} = \cF\left[ p, \E_{\rm UV}, \E_{\rm IR} , \lambda, \frac{N_q}{N_c^2} \right] \ ,
\end{eqnarray}
where $\cF$ is a functional. It is easy to check that at $\E_{\rm IR} = \E_{\rm cross} = \E_{\rm UV}$, $\cF > 0$ and parametrically large. This is based on the basic assumption that $N_q/N_c^2 \sim \O(1)$ and the 't Hooft coupling is large. 

Before leaving this section, let us note that in two-dimensional CFT, {\it i.e.}~when the bulk dual is an AdS$_3$, the effect of smeared light operators has been explored in \cite{Liu:2018iki, Hampapura:2018otw}.\footnote{We thank Shouvik Datta for pointing this out to us.} Generically the scrambling time is suppressed by the smeared factor in these case, as we have also observed in more involved systems in higher dimensions.

\section{Probe String in AdS$_3$}

Note that, in \cite{Banerjee:2018kwy, Banerjee:2018twd}, it was shown how an effective Schwarzian dynamics arises on an open string, that, subsequently, is responsible for a maximal chaos observed on the worldsheet Nambu-Goto theory. The The geometric background is AdS$_3 \times \cM_7$, where $\cM_7$ is a seven dimensional manifold whose explicit form is not relevant for us. The effective Schwarzian piece is essentially inherited from the embedding space of AdS$_3$. In the IR, however, the probe string becomes heavy and it is no longer legitimate to work in the probe limit. In \cite{Faedo:2014ana}, an exact solution was found by considering back-reaction of a smeared set of fundamental string sources, on the D$1$-D$5$ brane configuration, which gave rise to the AdS$_3$ supergravity background. 

Let us first look at the UV geometry, which has an AdS$_3$ piece. The metric is given by
\begin{eqnarray}
&& ds^2 = \left( \frac{u}{L}\right)^2 \left( -dt^2 + dx^2 \right) + L^2 \frac{du^2}{u^2} + L^2 ds_{S^3}^2 + \sqrt{\frac{Q_1}{v Q_5}} ds_{T^4}^2 \ , \\
&& L^4 = \frac{16 \pi^4}{V_4} \ell_s^4 g_s^2 \left( Q_1 Q_5 \right) \ , \quad v = \frac{V_4}{16 \pi^4 \ell_s^4} = {\rm fixed} \ ,
\end{eqnarray}
where, $S^3$ is a round three-sphere, $T^4$ is a four-torus, $V_4$ is the volume of the four-torus, and $Q_1$, $Q_5$ are the number of D$1$ and D$5$-branes, respectively. Clearly, for the holographic description, we need to take $Q_1, Q_5 \gg 1$ limit. Furthermore, since there are two underlying D-branes, we can subsequently define two 't Hooft couplings: $\lambda_1 = g_s Q_1$ and $\lambda_5 = g_s Q_5$. Thus, the string length can be related to the gauge couplings as:
\begin{eqnarray}
\ell_s^4 \sim \frac{L^4 } {\lambda_1 \lambda_5} \ .
\end{eqnarray}
Given this, the degrees of freedom on a string probe in the AdS$_3$ geometry therefore will be determined by 
\begin{eqnarray}
1 - \ell_s^{2} e^{\lambda_L t} = 1 -  \alpha \frac{1}{\sqrt{\lambda_1 \lambda_5}} e^{\lambda_L t} \quad \implies \quad t_{\rm scr} \sim \frac{1}{\lambda_L} \log \left( \lambda_1 \lambda_5 \right) \ . \label{scrUV}
\end{eqnarray}
In the above expression, $\alpha$ is a purely numerical constant.

The IR back-reacted geometry, in the string frame, is given by
\begin{eqnarray}
&& ds^2 = e^{\phi/2} \left[ \left( \frac{r}{L_{\rm IR}}\right)^{1/2} \left( - \left( \frac{r}{L_{\rm IR}}\right)^{4} dt^2 + \left( \frac{r}{L_{\rm IR}}\right)^{2} dx_1^2 + \left( \frac{L_{\rm IR}}{r}\right)^{2} dr^2 + \beta_x^2 Q^{2/5} dx_4^2 + \frac{L_{\rm IR}^2}{3} d\Omega_3^2\right) \right] \nonumber\\
&& L_{\rm IR} = \left( \frac{\beta_L}{Q}\right)^{1/5} L \ , \quad e^{\phi} = \beta_\phi \left(\frac{L_{\rm IR}}{L} \right)^4 \frac{r}{L_{\rm IR}} \ , \quad Q = \frac{2 \kappa^2}{2\pi \ell_s^2} \frac{L^{-1}}{2 V_{3}} N_q \ ,
\end{eqnarray}
where $\beta_{x,\phi, L}$ are purely numerical constants. To evaluate the effective tension of a fundamental probe string in the above background, we simply need to consider the $tt$ and the $rr$ components of the metric. In this IR geometry, a probe string effective tension is given by $\left( \ell_s  L \right)^{-2}$. Therefore, in this case, the scrambling time in the IR will also be given by the same result as in (\ref{scrUV}), except that the overall constant will be different. On the other hand, on a smeared set of probes the scrambling time picks up factors of the 't Hooft coupling as well and can be explicitly determined as we have done before.

\section{Conclusions}

In this work we have attempted an RG-perspective on dynamical correlation function. Within the modus operandi, we have left many interesting aspects untouched. For example, one can easily explore the role of a Kibble-Zurek type scaling relation between time and energy, on the correlators. Furthermore, given the nature of the Callan-Symanzik equation, exploring the solution space with characteristics appears a better suited option. We would like to address these issues in subsequent works. 

Our approach, in the QFT part, is mostly improvisational and not a systematic one. It would be quite interesting to find controllable examples where a first principle calculation can be carried out towards understanding the behaviour of OTOCs from an RG-perspective. While one is likely to run into subtleties in the perturbative treatment, as mentioned in \cite{Goykhman:2018iaz, DeSarker:2013abc}, it is certainly worth revisiting. We plan to address this in future.

The Holographic perspective, however, offers an interesting alternative in that this is no longer perturbative in the coupling constant. While this is true, a controllable quench-like dynamical description requires a full-fledged numerical exploration on how a black hole dynamically forms, if one is to avoid approximations. A possible avenue could be to focus on asymptotically AdS$_3$ geometries, in which still simpler examples may exist that capture the dynamics and allows us to relate it with a flow along the radial direction of the geometry. To our knowledge, this question is largely unexplored and we would also like to address it in future.

\acknowledgments

This work is supported by the Department of Atomic Energy, Govt. of India. We thank Diptarka Das, Shouvik Datta and Krishnendu Sengupta for various conversations related to this work. We specially thank Diptarka Das for many comments on the manuscript. A special thanks to Rakhi for her support and Dharitri for her constructive interruptions that made this work possible in this unusual time.

\appendix

\section{Stress-Tensor of Probes}

In this appendix, we will evaluate the stress-tensor of the probe sector, in various geometric backgrounds. We have considered two actions, one for the localized probes and one for smeared probes. These are given by
\begin{eqnarray}
&& S_{\rm probe} = - \frac{1}{2\pi \ell_s^2} \int d^2\sigma \sqrt{- {\rm det} \gamma} \int_{\Sigma_8} d\sigma^8 \ , \\
&& d\sigma_{\rm localized}^8 = \delta^{(8)}(y^i) dy^1 \wedge \ldots \wedge dy^8 \ , \quad d\sigma_{\rm smeared}^8 = \left( {\rm constant} \right) dy^1 \wedge \ldots \wedge dy^8 \ , \\
&& \gamma_{ab} = G_{\mu\nu} \left( \partial_a X^\mu\right) \left(\partial_b X^\nu\right) = G_{\mu\nu} e_a^\mu e_b^\nu \ . 
\end{eqnarray}
The corresponding stress-tensor is obtained as:
\begin{eqnarray}
T_{\mu\nu} = - \frac{2\kappa^2}{2\pi \ell_s^2} \frac{1}{\sqrt{- {\rm det} G}} \frac{\delta S}{\delta G^{\mu\nu}} =  - \frac{\kappa^2}{\pi \ell_s^2} \frac{1}{\sqrt{- {\rm det} G}} \frac{\delta S}{\delta \gamma^{ab}}\frac{\delta \gamma^{ab}}{\delta G^{\mu\nu}} \ .
\end{eqnarray}
Thus, for localized and for the smeared degrees of freedom, we get
\begin{eqnarray}
&& T_{\mu\nu}^{\rm loc} = - \frac{\kappa^2}{\pi \ell_s^2} \frac{1}{2} \frac{\sqrt{- {\rm det} \gamma} }{\sqrt{ - {\rm det} G}} \gamma_{ab} \frac{\delta \gamma^{ab}}{\delta G^{\mu\nu}} \ , \\
&& T_{\mu\nu}^{\rm smr} = - \frac{\kappa^2}{\pi \ell_s^2} \frac{1}{2} \frac{\sqrt{- {\rm det} \gamma} \sqrt{{\rm det} \gamma_{\rm smr}}}{\sqrt{ - {\rm det} G}} \gamma_{ab} \frac{\delta \gamma^{ab}}{\delta G^{\mu\nu}} \ .
\end{eqnarray}
In the above, we have denoted the weight of the smearing form by ${\rm det}\gamma_{\rm smr}$, which is proportional to the volume form along the directions transverse to the string worldsheet. Note that, this factor brings about the sole difference between the stress tensor of a localized probe and that of a smeared one. On the simple embedding that we have considered here, one obtains: $\left( {\rm det} G\right) = \left( {\rm det} \gamma\right) \left( {\rm det} \gamma_{\rm smr}\right)$.

Let us first catalogue the Einstein tensor for the IR-geometries in (\ref{backmet1})-(\ref{backmet4}), focussing on the $tt$-component only:
\begin{eqnarray}
&& E_{tt} \sim r^{26/3} \ , \quad p =1 \ , \label{em1} \\
&& E_{tt} \sim r^{10} \ , \quad p=2 \ , \label{em2} \\
&& E_{tt} \sim r^{14} \ , \quad p=3 \ , \label{em3} \\
&& E_{tt} \sim r^{2} \ , \quad p=4 \ . \label{em4}
\end{eqnarray}
In the above, the proportionality constant depends on order one numbers which we have ignored. Below, we present the result for various cases, separately:
\begin{eqnarray}
&& T_{\mu\nu}^{\rm loc} \sim r^{20/3} \ , \quad T_{\mu\nu}^{\rm smr} \sim r^{9} \ , \quad p =1 \ , \label{emt1} \\
&& T_{\mu\nu}^{\rm loc} \sim r^{5} \ , \quad T_{\mu\nu}^{\rm smr} \sim r^{11}  \ , \quad p =2 \ , \label{emt2} \\
&& T_{\mu\nu}^{\rm loc} \sim r^{2} \ , \quad T_{\mu\nu}^{\rm smr} \sim r^{17}  \ , \quad p =3 \ , \label{emt3} \\
&& T_{\mu\nu}^{\rm loc} \sim r^{-1} \ , \quad T_{\mu\nu}^{\rm smr} \sim r^{3} \ ,\quad p =4 \ . \label{emt4}
\end{eqnarray}
Now, comparison between (\ref{em1})-(\ref{em4}) and (\ref{emt1})-(\ref{emt4}) shows that for the smeared probes, the IR geometry receives no back-reaction at arbitrarily deep neck in the infrared; however, this is not the case for the localized probes. Thus, all energy scales appearing in the discussion here need to be larger than the scale where the localized probes begin back-reacting on the IR geometry itself.




\begin{thebibliography}{99}


\bibitem{Hayden:2007cs}
P.~Hayden and J.~Preskill,
``Black holes as mirrors: Quantum information in random subsystems,''
JHEP \textbf{09}, 120 (2007)
doi:10.1088/1126-6708/2007/09/120
[arXiv:0708.4025 [hep-th]].

\bibitem{Sekino:2008he}
Y.~Sekino and L.~Susskind,
``Fast Scramblers,''
JHEP \textbf{10}, 065 (2008)
doi:10.1088/1126-6708/2008/10/065
[arXiv:0808.2096 [hep-th]].

\bibitem{Lashkari:2011yi}
N.~Lashkari, D.~Stanford, M.~Hastings, T.~Osborne and P.~Hayden,
``Towards the Fast Scrambling Conjecture,''
JHEP \textbf{04}, 022 (2013)
doi:10.1007/JHEP04(2013)022
[arXiv:1111.6580 [hep-th]].

\bibitem{Yoshida:2017non}
B.~Yoshida and A.~Kitaev,
``Efficient decoding for the Hayden-Preskill protocol,''
[arXiv:1710.03363 [hep-th]].


\bibitem{Das:2016eao}
S.~R.~Das,
``Old and New Scaling Laws in Quantum Quench,''
PTEP \textbf{2016}, no.12, 12C107 (2016)
doi:10.1093/ptep/ptw146
[arXiv:1608.04407 [hep-th]].

\bibitem{Goykhman:2018iaz}
M.~Goykhman, T.~Shachar and M.~Smolkin,
``On quantum quenches at one loop,''
JHEP \textbf{01}, 022 (2019)
doi:10.1007/JHEP01(2019)022
[arXiv:1810.02258 [hep-th]].


\bibitem{DeSarker:2013abc}
S.~DeSarkar, R.~Sensarma and K.~Sengupta,
``A perturbative renormalization group approach to driven quantum systems,''
Journal of Physics: Condensed Matter \textbf{26}, 32 (2014)
doi: 10.1088/0953-8984/26/32/325602
[arXiv:1308.4689 [hep-th]].


\bibitem{Stanford:2015owe}
D.~Stanford,
``Many-body chaos at weak coupling,''
JHEP \textbf{10}, 009 (2016)
doi:10.1007/JHEP10(2016)009
[arXiv:1512.07687 [hep-th]].


\bibitem{Steinberg:2019uqb}
J.~Steinberg and B.~Swingle,
``Thermalization and chaos in QED$_{3}$,''
Phys. Rev. D \textbf{99} (2019) no.7, 076007
doi:10.1103/PhysRevD.99.076007
[arXiv:1901.04984 [cond-mat.str-el]].


\bibitem{Maldacena:2015waa}
J.~Maldacena, S.~H.~Shenker and D.~Stanford,
``A bound on chaos,''
JHEP \textbf{08}, 106 (2016)
doi:10.1007/JHEP08(2016)106
[arXiv:1503.01409 [hep-th]].


\bibitem{Itzhaki:1998dd}
N.~Itzhaki, J.~M.~Maldacena, J.~Sonnenschein and S.~Yankielowicz,
``Supergravity and the large N limit of theories with sixteen supercharges,''
Phys. Rev. D \textbf{58}, 046004 (1998)
doi:10.1103/PhysRevD.58.046004
[arXiv:hep-th/9802042 [hep-th]].


\bibitem{Shenker:2013pqa}
S.~H.~Shenker and D.~Stanford,
``Black holes and the butterfly effect,''
JHEP \textbf{03}, 067 (2014)
doi:10.1007/JHEP03(2014)067
[arXiv:1306.0622 [hep-th]].


\bibitem{Shenker:2013yza}
S.~H.~Shenker and D.~Stanford,
``Multiple Shocks,''
JHEP \textbf{12}, 046 (2014)
doi:10.1007/JHEP12(2014)046
[arXiv:1312.3296 [hep-th]].


\bibitem{Shenker:2014cwa}
S.~H.~Shenker and D.~Stanford,
``Stringy effects in scrambling,''
JHEP \textbf{05}, 132 (2015)
doi:10.1007/JHEP05(2015)132
[arXiv:1412.6087 [hep-th]].

\bibitem{Mateos:2006nu}
D.~Mateos, R.~C.~Myers and R.~M.~Thomson,
``Holographic phase transitions with fundamental matter,''
Phys. Rev. Lett. \textbf{97}, 091601 (2006)
doi:10.1103/PhysRevLett.97.091601
[arXiv:hep-th/0605046 [hep-th]].


\bibitem{Mateos:2007vn}
D.~Mateos, R.~C.~Myers and R.~M.~Thomson,
``Thermodynamics of the brane,''
JHEP \textbf{05}, 067 (2007)
doi:10.1088/1126-6708/2007/05/067
[arXiv:hep-th/0701132 [hep-th]].

\bibitem{Karch:2006bv}
A.~Karch and A.~O'Bannon,
``Chiral transition of N=4 super Yang-Mills with flavor on a 3-sphere,''
Phys. Rev. D \textbf{74}, 085033 (2006)
doi:10.1103/PhysRevD.74.085033
[arXiv:hep-th/0605120 [hep-th]].


\bibitem{Karch:2007pd}
A.~Karch and A.~O'Bannon,
``Metallic AdS/CFT,''
JHEP \textbf{09}, 024 (2007)
doi:10.1088/1126-6708/2007/09/024
[arXiv:0705.3870 [hep-th]].


\bibitem{Kundu:2013eba}
A.~Kundu and S.~Kundu,
``Steady-state Physics, Effective Temperature Dynamics in Holography,''
Phys. Rev. D \textbf{91}, no.4, 046004 (2015)
doi:10.1103/PhysRevD.91.046004
[arXiv:1307.6607 [hep-th]].


\bibitem{Alam:2012fw}
M.~Alam, V.~S.~Kaplunovsky and A.~Kundu,
``Chiral Symmetry Breaking and External Fields in the Kuperstein-Sonnenschein Model,''
JHEP \textbf{04}, 111 (2012)
doi:10.1007/JHEP04(2012)111
[arXiv:1202.3488 [hep-th]].


\bibitem{Johnson:2008vna}
C.~V.~Johnson and A.~Kundu,
``External Fields and Chiral Symmetry Breaking in the Sakai-Sugimoto Model,''
JHEP \textbf{12}, 053 (2008)
doi:10.1088/1126-6708/2008/12/053
[arXiv:0803.0038 [hep-th]].


\bibitem{Albash:2007bq}
T.~Albash, V.~G.~Filev, C.~V.~Johnson and A.~Kundu,
``Quarks in an external electric field in finite temperature large N gauge theory,''
JHEP \textbf{08}, 092 (2008)
doi:10.1088/1126-6708/2008/08/092
[arXiv:0709.1554 [hep-th]].


\bibitem{Albash:2006ew}
T.~Albash, V.~G.~Filev, C.~V.~Johnson and A.~Kundu,
``A Topology-changing phase transition and the dynamics of flavour,''
Phys. Rev. D \textbf{77}, 066004 (2008)
doi:10.1103/PhysRevD.77.066004
[arXiv:hep-th/0605088 [hep-th]].

\bibitem{Kim:2011qh}
K.~Y.~Kim, J.~P.~Shock and J.~Tarrio,
``The open string membrane paradigm with external electromagnetic fields,''
JHEP \textbf{06}, 017 (2011)
doi:10.1007/JHEP06(2011)017
[arXiv:1103.4581 [hep-th]].


\bibitem{Kundu:2019ull}
A.~Kundu,
``Steady States, Thermal Physics, and Holography,''
Adv. High Energy Phys. \textbf{2019}, 2635917 (2019)
doi:10.1155/2019/2635917


\bibitem{Kundu:2018sof}
A.~Kundu,
``Effective Thermal Physics in Holography: A Brief Review,''
[arXiv:1812.09447 [hep-th]].


\bibitem{Kundu:2015qda}
A.~Kundu,
``Effective Temperature in Steady-state Dynamics from Holography,''
JHEP \textbf{09}, 042 (2015)
doi:10.1007/JHEP09(2015)042
[arXiv:1507.00818 [hep-th]].


\bibitem{Liu:2018iki}
C.~Liu and D.~A.~Lowe,
``Notes on Scrambling in Conformal Field Theory,''
Phys. Rev. D \textbf{98}, no.12, 126013 (2018)
doi:10.1103/PhysRevD.98.126013
[arXiv:1808.09886 [hep-th]].


\bibitem{Hampapura:2018otw}
H.~R.~Hampapura, A.~Rolph and B.~Stoica,
``Scrambling in Two-Dimensional Conformal Field Theories with Light and Smeared Operators,''
Phys. Rev. D \textbf{99}, no.10, 106010 (2019)
doi:10.1103/PhysRevD.99.106010
[arXiv:1809.09651 [hep-th]].

\bibitem{Banerjee:2018kwy}
A.~Banerjee, A.~Kundu and R.~Poojary,
``Maximal Chaos from Strings, Branes and Schwarzian Action,''
JHEP \textbf{06}, 076 (2019)
doi:10.1007/JHEP06(2019)076
[arXiv:1811.04977 [hep-th]].


\bibitem{Banerjee:2018twd}
A.~Banerjee, A.~Kundu and R.~R.~Poojary,
``Strings, Branes, Schwarzian Action and Maximal Chaos,''
[arXiv:1809.02090 [hep-th]].

\bibitem{Murata:2017rbp}
K.~Murata,
``Fast scrambling in holographic Einstein-Podolsky-Rosen pair,''
JHEP \textbf{11}, 049 (2017)
doi:10.1007/JHEP11(2017)049
[arXiv:1708.09493 [hep-th]].


\bibitem{deBoer:2017xdk}
J.~de Boer, E.~Llabrés, J.~F.~Pedraza and D.~Vegh,
``Chaotic strings in AdS/CFT,''
Phys. Rev. Lett. \textbf{120}, no.20, 201604 (2018)
doi:10.1103/PhysRevLett.120.201604
[arXiv:1709.01052 [hep-th]].

\bibitem{Banerjee:2019vff}
A.~Banerjee, A.~Kundu and R.~R.~Poojary,
``Rotating Black Holes in AdS, Extremality and Chaos,''
[arXiv:1912.12996 [hep-th]].


\bibitem{Alishahiha:2016cjk}
M.~Alishahiha, A.~Davody, A.~Naseh and S.~F.~Taghavi,
``On Butterfly effect in Higher Derivative Gravities,''
JHEP \textbf{11}, 032 (2016)
doi:10.1007/JHEP11(2016)032
[arXiv:1610.02890 [hep-th]].



\bibitem{Faedo:2014ana}
A.~F.~Faedo, A.~Kundu, D.~Mateos and J.~Tarrio,
``(Super)Yang-Mills at Finite Heavy-Quark Density,''
JHEP \textbf{02}, 010 (2015)
doi:10.1007/JHEP02(2015)010
[arXiv:1410.4466 [hep-th]].
  


  
\end{thebibliography}
\end{document}